\documentclass{iopart}
\usepackage{iopams}
\usepackage{euscript,amssymb,epsfig}

\usepackage{amsfonts,latexsym}
\usepackage{euscript,amssymb}
\usepackage{epsfig}

\usepackage{graphicx}

\newcommand{\be}[1]{\begin{equation}\label{#1}}
\newcommand{\ee}{\end{equation}}
\newcommand{\ba}[1]{\begin{eqnarray}\label{#1}}
\newcommand{\ea}{\end{eqnarray}}
\newcommand{\rf}[1]{(\ref{#1})}
\newcommand{\nn}{\nonumber}

\renewcommand{\theequation}{\arabic{section}.\arabic{equation}}

\begin{document}

\title{Weak-field limit of Kaluza-Klein models with spherical compactification I: problematic aspects}

\author{Alexey Chopovsky, Maxim Eingorn and Alexander Zhuk}

\address{Astronomical Observatory and Department of
Theoretical Physics, Odessa National University, Street Dvoryanskaya 2, Odessa 65082, Ukraine}

\eads{\mailto{alexey.chopovsky@gmail.com}, \mailto{maxim.eingorn@gmail.com} and \mailto{ai$\_$zhuk2@rambler.ru}}


\begin{abstract} We investigate classical gravitational tests for the Kaluza-Klein model with spherical compactification of the internal two-dimensional space.
In the case of the absence of a multidimensional bare cosmological constant, the only matter which corresponds to the proposed metric ansatz is a perfect fluid with the
vacuum equation of state in the external space and the dust-like equation of state in the internal space. We perturb this background by a compact massive source with the
dust-like equation of state in both external and internal spaces (e.g., a point-like mass), and  obtain the metric coefficients in the weak-field approximation. It
enables to calculate the parameterized post-Newtonian parameter $\gamma$. We demonstrate that $\gamma=1/3$ which strongly contradicts the observations.
\end{abstract}

\pacs{04.25.Nx, 04.50.Cd, 04.80.Cc, 11.25.Mj}

\maketitle


\section{\label{sec:1}Introduction}

\setcounter{equation}{0}
Any physical theory is correct until it does not conflict with the experimental data.
Obviously, the Kaluza-Klein model is no exception to this rule.
There is a number of well-known gravitational experiments in the solar system, e.g., the deflection of light, the perihelion shift and the time delay of the radar echoes
(the Shapiro time-delay effect). In the weak-field limit, all these effects can be expressed via parameterized post-Newtonian (PPN) parameters $\beta$ and $\gamma$
\cite{Will,Straumann}. These parameters take different values in different gravitational theories. There are strict experimental restrictions on these parameters
\cite{Bertotti,Will2,JKh,CFPS}. The tightest constraint on $\gamma$ comes from the Shapiro time-delay experiment using the Cassini spacecraft: $\gamma -1 = (2.1\pm
2.3)\times 10^{-5}$. General Relativity is in good agreement with all gravitational experiments \cite{Landau}. Here, the PPN parameters $\beta=1$ and $\gamma=1$. The
Kaluza-Klein model should also be tested by the above-mentioned experiments.

In our previous papers \cite{EZ3,EZ4,EZ5} we have investigated this problem in the case of toroidal compactification of internal spaces. We have supposed that in the
absence of gravitating masses the metrics is a flat one. Gravitating compact objects (point-like masses or extended massive bodies) perturb this metrics, and we have
considered these perturbations in the weak-field approximation. First, we have shown that in the case of three-dimensional external/our space and dust-like equations of
state\footnote{\label{dust}Such equations of state take place in the external and internal spaces for a point-like mass at rest. For ordinary astrophysical extended
objects, e.g., for the Sun, it is also usually assumed that the energy density is much greater than pressure.} in the external and internal spaces, the PPN parameter
$\gamma = 1/(D-2)$, where $D$ is a total number of spatial dimensions. Obviously, $D=3$  (i.e. the General Relativity case) is the only value which does not contradict
the observations \cite{EZ3}. Second, in papers \cite{EZ4,EZ5}, we have investigated the exact soliton solutions. In these solutions a gravitating source is uniformly
smeared over the internal space and the non-relativistic gravitational potential exactly coincides with the Newtonian one. Here, we have found a class of solutions which
are indistinguishable from General Relativity. We have called such solutions latent solitons. Black strings and black branes belong to this class. They have the
dust-like equation of state $p_0=0$ in the external space and the relativistic equation of state $p_1=-\varepsilon/2$ in the internal space. It is known (see
\cite{EZ5,Zhuk}) that in the case of three-dimensional external space with a dust-like perfect fluid, this combination of equations of state in the external and internal
spaces does not spoil the internal space stabilization. Moreover, we have shown also that the number $d_0=3$ of the external dimensions is unique. Therefore, there is no
problem for black strings and black branes to satisfy the gravitational experiments in the solar system at the same level of accuracy as General Relativity. However, the
main problem with the black strings/branes is to find a physically reasonable mechanism which can explain how the ordinary particles forming the astrophysical objects
can acquire rather specific equations of state $p_i=-\varepsilon/2$ (tension!) in the internal spaces. Thus, in the case of toroidal compactification, on the one hand we
arrive at the contradiction with the experimental data for the physically reasonable gravitating source in the form of a point-like mass, on the other hand we have no
problem with the experiments for black strings/branes but arrive at very strange equation of state in the internal space. How common is this problem for the Kaluza-Klein
models?

To understand it, in the present paper we investigate a model with spherical compactification of the internal space. Therefore, in contrast to the previous case the
background metrics is not flat but has a topology $\mathbb{R}\times\mathbb{R}^3\times S^2$. To make the internal space curved, we must introduce a background matter. We
show that in the case of the absence of a bare six-dimensional cosmological constant, the only matter which corresponds to this background metrics is the one which
simulates a perfect fluid with the vacuum equation of state in the flat external space and the dust-like equation of state in the curved internal space. To get the PPN
parameters in this model, we perturb the background metrics and matter by a compact gravitating massive object with the dust-like equation of state in the external and
internal spaces (e.g., a point-like mass). Our investigation shows that we arrive at the same conclusions as in the case of the toroidal compactification, e.g., the PPN
parameter $\gamma =1/3$ which exactly coincides with the formula $\gamma =1/(D-2)$ for $D=5$. Obviously, this value contradicts the observations.

The paper is organized as follows. In section 2 we get the background matter corresponding to the background metric ansatz. Then we perturb this background by the
massive compact object with the dust-like equations of state and obtain the perturbed metric coefficients. It gives us a possibility to calculate the PPN parameter
$\gamma$ in section 3. The main results are summarized in section 4. In appendixes A and B we present formulae for the components of the Ricci tensor   and, with the
help of them, investigate the relations between the perturbed metric coefficients.

\section{\label{sec:2}Background solution and perturbations}

\setcounter{equation}{0}

To start with, let us consider a factorizable six-dimensional static background metrics
\be{2.1}
 ds^2=c^2 dt^2-dx^2-dy^2-dz^2-a^2(d\xi^2+\sin^2\xi d\eta^2)
\ee
which is defined on a product manifold $M = M_4\times M_2$. $M_4$ describes external four-dimensional flat space-time and $M_2$ corresponds to the two-dimensional
internal space which is a sphere with the radius (the internal space scale factor) $a$. Now, we want to define the form of the energy-momentum tensor of matter which
corresponds to this geometry. In contrast to the models in \cite{EZ3,EZ4,EZ5} with toroidal compactification where the external and internal background metrics are flat
and there is no need for the matter to create such a flat background, in the present paper, we need such a bare matter to make a curved internal space. Obviously, this
form is defined by the Einstein equation
\be{2.2}
\kappa T_{ik} = R_{ik}-\frac{1}{2} R g_{ik}\, ,
\ee
where $\kappa \equiv 2S_5\tilde G_{6}/c^4$. Here, $S_5=2\pi^{5/2}/\Gamma (5/2)=8\pi^2/3$ is the total solid angle (the surface area of the four-dimensional sphere of a
unit radius) and $\tilde G_{6}$ is the gravitational constant in the six-dimensional space-time.

As it can be easily seen from appendix A, the only nonzero components of the Ricci tensor for the metrics \rf{2.1} are $R_{44}=1$ and $R_{55}=\sin^2 \xi$, and the scalar
curvature is $R=-2/ a^2$. Therefore, the energy-momentum tensor satisfies the following condition:
\be{2.3}
T_{ik}=\left\{
\begin{array}{cc}
\left( 1/\left(\kappa a^2\right) \right) g_{ik} & \mbox{for   } \, i,k=0,...,3;\\
\\
0 & \mbox{for   } \, i, k=4,5.
\end{array} \right . \quad
\ee
Clearly, such matter can be simulated by a perfect fluid with the vacuum equation of state in the external space and the dust-like equation of state in the internal
space. It is convenient to introduce the following notation: $\Lambda_4\equiv 1/(\kappa a^2)$. Because of the flatness of the external space-time in \rf{2.1}, an
effective four-dimensional cosmological constant \cite{exci,GSZ} $\Lambda_{(4)eff}$ should be equal to zero. Indeed, it is not difficult to verify that it takes place in
the considered case.

Now, we want to perturb this background by a point-like mass. It is well known that a point-like massive source is a good physical approximation in four-dimensional
space-time to calculate the classical gravitational tests \cite{Landau}. These calculations show that General Relativity is in good agreement with observational data. We
intend to get the corresponding formulae in the case of the six-dimensional background metrics \rf{2.1} in the presence of the background matter \rf{2.3}, and to compare
the obtained results with the known observational data. To perform it, we perturb our background ansatz by a static point-like massive source with non-relativistic rest
mass density $\epsilon \rho ({\bf r}_5)$. We introduce an infinitesimal prefactor $\epsilon$ to keep during calculations the corresponding orders of perturbations.
At the end of calculations this parameter should be set equal to unity. It is worth noting that we take into account the point-like nature of the matter source only for
the calculation of the non-relativistic gravitational potential (see the next section). In the present section we do not specify the concrete form of $\rho ({\bf r}_5)$.
In this general case, we assume that the dust-like (i.e. pressure is much less than energy density) massive source with the rest mass density  $\rho$ represents a static
compact astrophysical object.
There are two separate cases. In the first case, the matter source is uniformly smeared over the internal space. Here, multidimensional $\rho$ and three-dimensional
$\rho_3$ rest mass densities are connected as follows: $\rho=\rho_3({\bf r}_3)/(4\pi a^2)$. In the case of a point-like mass $m$, $\rho_3(r_3)=m\delta (\bf {r}_3)$,
where $r_3=|{\bf r}_3|=\sqrt{x^2+y^2+z^2}$. In the second case (without smearing), the rest mass density is a function of all five spatial coordinates. In the present
paper we consider mainly the latter case.

For the perturbed metrics we choose the metric ansatz in the form
\be{2.4}
ds^2=Ac^2dt^2+Bdx^2+Cdy^2+Ddz^2+Ed\xi^2+Fd\eta^2\, ,
\ee
where the metric coefficients $A,B,C,D,E$ and $F$ are functions of all spatial coordinates, e.g., $A=A({\bf r}_5)$. We also suppose that, up to corrections of the first
order in $\epsilon$, the metric coefficients read
\ba{2.5}
A&\approx &A^{0}+\epsilon A^{1}({\bf r}_5),\quad B \approx B^{0}+\epsilon B^{1}({\bf r}_5),\quad C \approx C^{0}+\epsilon C^{1}({\bf r}_5),\nn\\
D& \approx &D^{0}+\epsilon D^{1}({\bf r}_5),\quad E \approx E^{0}+\epsilon E^{1}({\bf r}_5),\quad F \approx F^{0}+\epsilon F^{1}({\bf r}_5)\, , \ea
where the metric coefficients $A^0,B^0,C^0,D^0,E^0$ and $F^0$ are defined by the background metrics \rf{2.1}:
\be{2.7}
A^{0}=1,\quad B^{0}=C^{0}=D^{0}=-1,\quad E^{0}=-a^2,\quad F^{0}=E^{0}\sin^2\xi\, .
\ee

We suppose that the perturbed metrics preserves its diagonal form. Obviously, the off-diagonal coefficients $g_{0\alpha}, \, \alpha =1,\ldots ,5$, are absent for the
static metrics. It is also clear that in the case of uniformly smeared (over the internal space) perturbation, all metric coefficients depend only on $x,y,z$ (see, e.g.,
\cite{EZ2}), and the metric structure of the internal space does not change, i.e. $F=E\sin ^2\xi$. It is not difficult to show, that in this case the spatial part of the
external metrics can be diagonalized by coordinate transformations. Moreover, if we additionally assume the spherical symmetry of the perturbation with respect to the
external space, then all metric coefficients depend on $r_3$ and $B(r_3)=C(r_3)=D(r_3)$. Taking into account all these arguments, we suppose that the diagonal form is
preserved also for an arbitrary distribution $\epsilon \rho (\bf r_5)$, and we show below that Einstein's equations have solution for the given metric ansatz.

Now, for the metric ansatz \rf{2.4}, we want to solve the Einstein equation \rf{2.2} which we rewrite in the form
\be{2.8}
R_{ik}=\kappa\left( T_{ik}-\frac{1}{4}Tg_{ik} \right)\, .
\ee
The energy-momentum tensor consists of two parts:
\be{2.9}
T_{ki}=\tilde{T}_{ki}+\hat{T}_{ki}\, .
\ee
Here, $\tilde{T}_{ki}$ is the energy-momentum tensor of the perturbed background matter \rf{2.3} and $\hat{T}_{ki}$ is the energy-momentum tensor of the perturbation. In
the non-relativistic approximation the only nonzero component of the latter tensor is $\hat{T}^0_{0}\approx \epsilon \rho ({\bf r_5}) c^2$ and up to linear in
$\epsilon$ terms $\hat{T}_{00}\approx \epsilon \rho ({\bf r_5}) c^2$. Concerning the energy momentum tensor of the background matter, we suppose that perturbation
does not change the equations of state in the external and internal spaces. For example, if we had dust in the internal space before the perturbation, the same equation
of state should be preserved after the perturbation. The vacuum equation of state in the external space should be also preserved. Here, the perturbation results in the
appearance of a small fluctuation: $\Lambda_4 \; \to \; \Lambda_4 + \epsilon \Lambda_4^{(1)}$. Therefore, up to the first order correction terms, the nonzero
components of the energy-momentum tensor read
\ba{2.10}
T_{00}&\approx&\frac{1}{\kappa a^2}+\epsilon\left(\frac{1}{\kappa a^2}A^1+\rho c^2+\Lambda_4^{(1)}\right)\, , \\
\label{2.11}T_{ii}&\approx&-\frac{1}{\kappa a^2}+\epsilon\left(\frac{1}{\kappa a^2}B_i-\Lambda_4^{(1)}\right)\, ,\quad i=1,2,3\, ,
\ea
where $B_1\equiv B^1,B_2\equiv C^1$ and $B_3\equiv D^1$.
The trace of total energy-momentum tensor is
\be{2.12}
T=T_{i}^{i}\approx\frac{4}{\kappa a^2}+\epsilon\left(\rho c^2+4\Lambda_4^{(1)}\right)\, .
\ee
Therefore, the diagonal components of the Einstein equation \rf{2.8} up to linear in $\epsilon$ terms read:
\ba{2.13}
R_{00} &\approx& \epsilon\frac{3}{4}\kappa\rho c^2\, ,\\
\label{2.14}R_{11}&=&R_{22}=R_{33} \approx  \epsilon\frac{1}{4}\kappa\rho  c^2\, ,\\
\label{2.15} R_{44}& \approx & \, 1 -\epsilon\left(\frac{E^1}{a^2}- \kappa\Lambda_4^{(1)}a^2-\frac{a^2}{4}\kappa\rho  c^2\right)\, ,\\
\label{2.16} R_{55} &\approx & \sin^2\xi -\epsilon\left(\frac{F^1}{a^2}- \kappa\Lambda_4^{(1)}a^2\sin^2\xi-\frac{a^2\sin^2\xi}{4}\kappa\rho  c^2\right)\, .
\ea
Taking into account \rf{a5}, we can write the $00$-component as follows:
\be{2.17}
\triangle_3 A^1 +\frac{1}{a^2}\triangle_{\xi\eta}A^1=\frac{3}{2}\kappa\rho  c^2\, .
\ee
All off-diagonal components of the Einstein equation \rf{2.8} are equal to zero: $R_{ik}=0\, , i\ne k$. Therefore, we can use the results of appendix B. Then, all three
components $11$, $22$ and $33$ are reduced to one equation
\be{2.18}
\triangle_3 B^1 +\frac{1}{a^2}\triangle_{\xi\eta}B^1=\frac{1}{2}\kappa\rho  c^2\, ,
\ee
where we use expressions \rf{a7}-\rf{a9} and relations \rf{b1}. Hence, the metric coefficients $A^1$ and $B^1$ are related as follows: $A^1 =3B^1$, i.e. it is the case 2
of appendix B. It is not difficult to verify with the help of \rf{b13}, \rf{b15} and \rf{b16} that $44$ and $55$ components are reduced to one equation
\be{2.19}
\triangle_3 E^1 +\frac{1}{a^2}\triangle_{\xi\eta}E^1=
\frac{1}{2}\kappa\rho c^2a^2+2\kappa\Lambda_4^{(1)}a^2-\frac{2}{a^2}E^1\, .
\ee
From the system of equations \rf{2.17}-\rf{2.19} and the relation \rf{b9} (which is also valid in the case 2) we can conclude that
\be{2.20} \kappa\Lambda_4^{(1)}=\frac{E^1}{a^4} \ee
and equation \rf{2.19} reads
\be{2.21}
\triangle_3 E^1 +\frac{1}{a^2}\triangle_{\xi\eta}E^1=\frac{a^2}{2}\kappa\rho  c^2\, .
\ee


\section{\label{sec:3}Parameterized post-Newtonian parameter $\gamma$: a contradiction to observations}

\setcounter{equation}{0}

Now let us solve the equation \rf{2.17}. Denoting $A^1\equiv 2\varphi/c^2$, we obtain the Poisson equation:
\be{4.1} \triangle_3 \varphi +\frac{1}{a^2}\triangle_{\xi\eta}\varphi = S_5G_6\rho({\bf r}_5)\, , \ee
where $G_6=3\tilde G_6/2$. In the case of toroidal compactification, we have $G_{\mathcal{D}}=[2(D-2)/(D-1)]\tilde G_{\mathcal{D}}$ where ${\mathcal{D}}=D+1$ is an
arbitrary number of space-time dimensions \cite{EZ3}. So, our particular formula follows from this general relation for ${\mathcal{D}}=6$. It is not difficult to verify
that in the case of a point-like mass $m$ with $\rho = m\delta ({\bf r}_5)$, the equation \rf{4.1} has the following solution (see also \cite{KehSf}):
\be{4.2} \fl \varphi =-\frac{S_5G_6}{4\pi a^2} \frac{m}{r_3}\sum_{l=0}^{+\infty}\sum_{m=-l}^{ l}Y^*_{lm}(\xi_0,\eta_0)\, Y_{lm}(\xi, \eta)\,
\exp\left(-\frac{\sqrt{l(l+1)}}{a}\,\, r_3\right)\, , \ee
where $\xi_0,\eta_0$ denote the position of the source on the two-dimensional sphere (without loss of generality we can choose $\xi_0=0$, $\eta_0 =0$) and $Y_{lm}$ are
Laplace's spherical harmonics. Obviously, in the limit $r_3\to +\infty$, the non-relativistic gravitational potential should coincide with the Newtonian one:
\be{4.3}
\varphi(r_3\rightarrow +\infty)\rightarrow-\frac{G_N m}{r_3}\, ,
\ee
where $G_N$ is the Newtonian gravitational constant. Taking into account that the zero Kaluza-Klein mode $l=0$, $m=0$ gives the main contribution in \rf{4.2} for $r_3\to
+\infty$ and that $Y_{00}=1/\sqrt{4\pi}$, we get the relation between six-dimensional and Newtonian gravitational constants
\be{4.4} \frac{S_5 G_6}{4\pi a^2} = 4\pi G_N \ee
which exactly coincides with the corresponding relation in the case of toroidal compactification for ${\mathcal{D}}=6$ and the internal space volume $V_2=4\pi a^2$ (see,
e.g., the equation (20) in \cite{EZ4}).

Therefore, the perturbation $A^1$ of the $00$ metric coefficient reads
\be{4.5}
\fl A^1=-4\pi \frac{r_g}{r_3}\sum_{l=0}^{+\infty}\sum\limits_{m=-l}^l\,Y^*_{lm}(\xi_0,\eta_0)\,Y_{lm}(\xi, \eta)\exp \left( -\frac{\sqrt {l(l+1)}}{a}\,r_3\right)\, ,
\ee
where the gravitational radius $r_g= 2G_N m/c^2$. Perturbations of other metric coefficients can be found with the help of relations \rf{b13}. Obviously, the radius of
the astrophysical objects, such as the Sun, is much larger than the compactification scale of the internal space: $R_3\gg a$. Then, for $r_3 \gtrsim R_3$ we can limit
ourselves to the zero mode in \rf{4.5}. Therefore, at these distances the metrics \rf{2.4} reads
\ba{4.6}
ds^2&\approx& \left(1-\frac{r_g}{r_3}\right)c^2dt^2-\left(1+\frac13\frac{r_g}{r_3}\right)
\left(dx^2+dy^2+dz^2\right)\nn \\
&-&a^2\left(1+\frac13\frac{r_g}{r_3}\right)\left(d\xi^2+\sin^2\xi d\eta^2\right)\, .
\ea
In the case of the matter source which is uniformly smeared over the internal space (a particular example of the case 2 in appendix B), we have $\rho ({\bf r}_5)=m\delta
({\bf r}_3)/(4\pi a^2)$, and the equation \rf{4.1} is reduced to the ordinary three-dimensional Poisson equation $\triangle_3 \varphi = 4\pi G_N m\delta ({\bf r}_3)$
with the solution $\varphi = -G_N m/r_3\; \to \; A^1=-r_g/r_3$.

It can be easily seen from the expression \rf{4.6} (see \cite{Will,Straumann}), that the parameterized post-Newtonian (PPN) parameter $\gamma$ reads
\be{4.7}
\gamma = \frac13\, .
\ee
The tightest constraint on $\gamma$ comes from  the Shapiro time-delay experiment using the Cassini spacecraft:
$\gamma-1 =(2.1\pm 2.3)\times 10^{-5}$ \cite{Bertotti,Will2,JKh,CFPS}.
Obviously, the PPN parameter $\gamma$ \rf{4.7} does not satisfy this restriction.

It is worth noting that in the case of toroidal compactification we get $\gamma=1/(D-2)$ \cite{EZ3} which results exactly in our formula \rf{4.7} for $D=5$. Therefore,
spherical compactification with considered background matter \rf{2.3} does not save the situation with the point-like massive source. Similar to the case of toroidal
compactification, we also come to a contradiction to the observations. The metrics \rf{b14} indicates that the same conclusion must also occur in the case of compact
astrophysical objects (not necessarily point-like) with dust-like equations of state in the external and internal spaces and an arbitrary distribution $\rho ({\bf
r}_5)$. It can be easily seen that in this general case the ratio $B^1/A^1 = 1/3$ for $\forall \; \varphi$. However, to satisfy the experimental constraints, this ratio
should be very close to 1, namely $1 + (2.1\pm 2.3)\times 10^{-5}$. Therefore, there is a common feature among the model with toroidal compactification and the present
model with spherical compactification which leads to a contradiction with the observations.

\section{Conclusion}

In our paper we investigated classical gravitational tests for the Kaluza-Klein model with spherical compactification of the internal space. The external spacetime is
flat. We supposed that a multidimensional bare cosmological constant is absent. In this case, the only matter which corresponds to proposed metric ansatz is the one
which can be simulated by a perfect fluid with the vacuum equation of state in the external space and the dust-like equation of state in the internal space. We perturbed
this background by a compact massive source with the dust-like equation of state in both spaces. In the weak-field limit, the perturbed metric coefficients were found as
a solution of the system of Einstein equations. It enabled to calculate the PPN parameter $\gamma$. We found that for our model $\gamma=1/3$ which strongly contradicts
the observations. The similar situation takes place for models with toroidal compactification. We think that it happens because in both of these types of models the
internal spaces are not stabilized. The second part of our research will be devoted to solving this problem. We shall see that our guess is correct and in the case of
stabilized internal spaces considered models can be in agreement with observations.

\ack We want to thank Prof. E.K. Loginov for useful discussions. This work was supported in part by the "Cosmomicrophysics" programme of the Physics and Astronomy
Division of the National Academy of Sciences of Ukraine.


\section{Appendix A: Components of the Ricci tensor}
\renewcommand{\theequation}{A\arabic{equation}}
\setcounter{equation}{0}

In this appendix we consider the six-dimensional space-time metrics of the form of \rf{2.4}:
$$
ds^2=Ac^2dt^2+Bdx^2+Cdy^2+Ddz^2+Ed\xi^2+Fd\eta^2\, ,
$$
where the metric coefficients $A,B,C,D,E$ and $F$ satisfy the decomposition \rf{2.5}. Now, we define the corresponding components of the Ricci tensor up to linear in
$\epsilon$ terms.

\subsection{Diagonal components}

\ba{a5} \fl &{}&R_{00}\nn\\
\fl &=&\frac{B_t}{4c^2B}\left(-\frac{2B_{tt}}{B_t}+\frac{A_t}{A}+\frac{B_t}{B}\right)+\frac{A_x}{4B}
\left(-\frac{2A_{xx}}{A_x}+\frac{B_x}{B}-\frac{C_x}{C}-\frac{D_x}{D}-\frac{E_x}{E}-\frac{F_x}{F}+\frac{A_x}{A}\right)\nn
\\
\fl &+&\frac{C_t}{4c^2C}\left(-\frac{2C_{tt}}{C_t}+\frac{A_t}{A}+\frac{C_t}{C}\right)+\frac{A_y}{4C}
\left(-\frac{2A_{yy}}{A_y}+\frac{C_y}{C}-\frac{B_y}{B}-\frac{D_y}{D}-\frac{E_y}{E}-\frac{F_y}{F}+\frac{A_y}{A}\right)\nn
\\
\fl &+&\frac{D_t}{4c^2D}\left(-\frac{2D_{tt}}{D_t}+\frac{A_t}{A}+\frac{D_t}{D}\right)+\frac{A_z}{4D}
\left(-\frac{2A_{zz}}{A_z}+\frac{D_z}{D}-\frac{B_z}{B}-\frac{C_z}{C}-\frac{E_z}{E}-\frac{F_z}{F}+\frac{A_z}{A}\right)\nn
\\
\fl &+&\frac{E_t}{4c^2E}\left(-\frac{2E_{tt}}{E_t}+\frac{A_t}{A}+\frac{E_t}{E}\right)+\frac{A_{\xi}}{4E}
\left(-\frac{2A_{\xi\xi}}{A_{\xi}}+\frac{E_{\xi}}{E}-\frac{B_{\xi}}{B}-\frac{C_{\xi}}{C}-\frac{D_{\xi}}{D}-\frac{F_{\xi}}{F}+\frac{A_{\xi}}{A}\right)\nn
\\
\fl &+&\frac{F_t}{4c^2F}\left(-\frac{2F_{tt}}{F_t}+\frac{A_t}{A}+\frac{F_t}{F}\right)+\frac{A_{\eta}}{4F}
\left(-\frac{2A_{\eta\eta}}{A_{\eta}}+\frac{F_{\eta}}{F}-\frac{B_{\eta}}{B}-\frac{C_{\eta}}{C}-\frac{D_{\eta}}{D}-\frac{E_{\eta}}{E}+\frac{A_{\eta}}{A}\right)\nn\\
\fl &\approx& \frac{\epsilon}{2}\left[\triangle_3 A^1 +\frac{1}{a^2}\triangle_{\xi\eta}A^1\right]\, .
\ea
Here, indexes denote the corresponding partial derivatives (e.g., $A_{x}\equiv \partial A/\partial x$) and we introduce the Laplace operators:
\be{a6} \fl \triangle_3\equiv \frac{\partial^2}{\partial x^2} + \frac{\partial^2}{\partial y^2} + \frac{\partial^2}{\partial z^2}\, ,\quad \triangle_{\xi\eta}\equiv
\frac{\partial^2}{\partial \xi^2} + \frac{\cos \xi}{\sin \xi}\frac{\partial}{\partial\xi} + \frac{1}{\sin^2\xi}\frac{\partial^2}{\partial \eta^2}\, . \ee
For $R_{11}$, $R_{22}$ and $R_{33}$ we obtain respectively:
\ba{a7} \fl &{}& R_{11}\nn\\
\fl &=& \frac{A_x}{4A}\left(-\frac{2A_{xx}}{A_x}+\frac{B_x}{B}+\frac{A_x}{A}\right)+\frac{B_t}{4c^2A}
\left(-\frac{2B_{tt}}{B_t}+\frac{A_t}{A}-\frac{C_t}{C}-\frac{D_t}{D}-\frac{E_t}{E}-\frac{F_t}{F}+\frac{B_t}{B}\right)\nn
\\
\fl &+&\frac{C_x}{4C}\left(-\frac{2C_{xx}}{C_x}+\frac{B_x}{B}+\frac{C_x}{C}\right)+\frac{B_y}{4C}
\left(-\frac{2B_{yy}}{B_y}+\frac{C_y}{C}-\frac{A_y}{A}-\frac{D_y}{D}-\frac{E_y}{E}-\frac{F_y}{F}+\frac{B_y}{B}\right)\nn
\\
\fl &+&\frac{D_x}{4D}\left(-\frac{2D_{xx}}{D_x}+\frac{B_x}{B}+\frac{D_x}{D}\right)+\frac{B_z}{4D}
\left(-\frac{2B_{zz}}{B_z}+\frac{D_z}{D}-\frac{A_z}{A}
-\frac{C_z}{C}-\frac{E_z}{E}-\frac{F_z}{F}+\frac{B_z}{B}\right)\nn
\\
\fl &+&\frac{E_x}{4E}\left(-\frac{2E_{xx}}{E_x}+\frac{B_x}{B}+\frac{E_x}{E}\right)+\frac{B_{\xi}}{4E}
\left(-\frac{2B_{\xi\xi}}{B_{\xi}}+\frac{E_{\xi}}{E}-\frac{A_{\xi}}{A}-\frac{C_{\xi}}{C}-\frac{D_{\xi}}{D}-\frac{F_{\xi}}{F}+\frac{B_{\xi}}{B}\right)\nn\\
\fl &+&\frac{F_x}{4F}\left(-\frac{2F_{xx}}{F_x}+\frac{B_x}{B}+\frac{F_x}{F}\right)+\frac{B_{\eta}}{4F}
\left(-\frac{2B_{\eta\eta}}{B_{\eta}}+\frac{F_{\eta}}{F}-\frac{A_{\eta}}{A}-\frac{C_{\eta}}{C} -\frac{D_{\eta}}{D}-\frac{E_{\eta}}{E}+\frac{B_{\eta}}{B}\right)\nn
\\
\fl &\approx& -\frac{\epsilon}{2}\left[-\triangle_3 B^1
+\left(A^1+B^1-C^1-D^1-\frac{E^1}{a^2}-\frac{F^1}{a^2\sin^2\xi}\right)_{xx}-\frac{1}{a^2}\triangle_{\xi\eta}B^1\right]\, , \ea
\ba{a8} \fl &{}& R_{22}\nn\\
\fl &=&\frac{A_y}{4A}\left(-\frac{2A_{yy}}{A_y}+\frac{C_y}{C}+\frac{A_y}{A}\right)+\frac{C_t}{4c^2A}
\left(-\frac{2C_{tt}}{C_t}+\frac{A_t}{A}-\frac{B_t}{B}-\frac{D_t}{D}-\frac{E_t}{E}-\frac{F_t}{F}+\frac{C_t}{C}\right)\nn
\\
\fl &+&\frac{B_y}{4B}\left(-\frac{2B_{yy}}{B_y}+\frac{C_y}{C}+\frac{B_y}{B}\right)+\frac{C_x}{4B}
\left(-\frac{2C_{xx}}{C_x}+\frac{B_x}{B}-\frac{A_x}{A}-\frac{D_x}{D}-\frac{E_x}{E}-\frac{F_x}{F}+\frac{C_x}{C}\right)\nn
\\
\fl &+&\frac{D_y}{4D}\left(-\frac{2D_{yy}}{D_y}+\frac{C_y}{C}+\frac{D_y}{D}\right)+\frac{C_z}{4D}
\left(-\frac{2C_{zz}}{C_z}+\frac{D_z}{D}-\frac{A_z}{A}-\frac{B_z}{B}-\frac{E_z}{E}-\frac{F_z}{F}+\frac{C_z}{C}\right)\nn
\\
\fl &+&\frac{E_y}{4E}\left(-\frac{2E_{yy}}{E_y}+\frac{C_y}{C}+\frac{E_y}{E}\right)+\frac{C_{\xi}}{4E}
\left(-\frac{2C_{\xi\xi}}{C_{\xi}}+\frac{E_{\xi}}{E}-\frac{A_{\xi}}{A}-\frac{B_{\xi}}{B}-\frac{D_{\xi}}{D}-\frac{F_{\xi}}{F}+\frac{C_{\xi}}{C}\right)\nn
\\
\fl &+&\frac{F_y}{4F}\left(-\frac{2F_{yy}}{F_y}+\frac{C_y}{C}+\frac{F_y}{F}\right)+\frac{C_{\eta}}{4F}
\left(-\frac{2C_{\eta\eta}}{C_{\eta}}+\frac{F_{\eta}}{F}-\frac{A_{\eta}}{A}-\frac{B_{\eta}}{B}-\frac{D_{\eta}}{D}-\frac{E_{\eta}}{E}+
\frac{C_{\eta}}{C}\right)\nn
\\
\fl &\approx& -\frac{\epsilon}{2}\left[-\triangle_3 C^1
+\left(A^1+C^1-B^1-D^1-\frac{E^1}{a^2}-\frac{F^1}{a^2\sin^2\xi}\right)_{yy}-\frac{1}{a^2}\triangle_{\xi\eta}C^1\right]\, , \ea
\ba{a9} \fl &{}& R_{33}\nn\\
\fl &=& \frac{A_z}{4A}\left(-\frac{2A_{zz}}{A_z}+\frac{D_z}{D}+\frac{A_z}{A}\right)+\frac{D_t}{4c^2A}
\left(-\frac{2D_{tt}}{D_t}+\frac{A_t}{A}-\frac{B_t}{B}-\frac{C_t}{C}-\frac{E_t}{E}-\frac{F_t}{F}+\frac{D_t}{D}\right)\nn
\\
\fl &+&\frac{B_z}{4B}\left(-\frac{2B_{zz}}{B_z}+\frac{D_z}{D}+\frac{B_z}{B}\right)+\frac{D_x}{4B}
\left(-\frac{2D_{xx}}{D_x}+\frac{B_x}{B}-\frac{A_x}{A}-\frac{C_x}{C}-\frac{E_x}{E}-\frac{F_x}{F}+\frac{D_x}{D}\right)\nn
\\
\fl &+&\frac{C_z}{4C}\left(-\frac{2C_{zz}}{C_z}+\frac{D_z}{D}+\frac{C_z}{C}\right)+\frac{D_y}{4C}
\left(-\frac{2D_{yy}}{D_y}+\frac{C_y}{C}-\frac{A_y}{A}-\frac{B_y}{B}-\frac{E_y}{E}-\frac{F_y}{F}+\frac{D_y}{D}\right)\nn
\\
\fl &+&\frac{E_z}{4E}\left(-\frac{2E_{zz}}{E_z}+\frac{D_z}{D}+\frac{E_z}{E}\right)+\frac{D_{\xi}}{4E}
\left(-\frac{2D_{\xi\xi}}{D_{\xi}}+\frac{E_{\xi}}{E}-\frac{A_{\xi}}{A}-\frac{B_{\xi}}{B}-\frac{C_{\xi}}{C}-\frac{F_{\xi}}{F}+\frac{D_{\xi}}{D}\right)\nn
\\
\fl &+&\frac{F_z}{4F}\left(-\frac{2F_{zz}}{F_z}+\frac{D_z}{D}+\frac{F_z}{F}\right)+\frac{D_{\eta}}{4F}
\left(-\frac{2D_{\eta\eta}}{D_{\eta}}+\frac{F_{\eta}}{F}-\frac{A_{\eta}}{A}-\frac{B_{\eta}}{B}-\frac{C_{\eta}}{C}-
\frac{E_{\eta}}{E}+\frac{D_{\eta}}{D}\right)\nn
\\
\fl &\approx& -\frac{\epsilon}{2}\left[-\triangle_3 D^1
+\left(A^1+D^1-B^1-C^1-\frac{E^1}{a^2}-\frac{F^1}{a^2\sin^2\xi}\right)_{zz}-\frac{1}{a^2}\triangle_{\xi\eta}D^1\right]\, . \ea
The components $R_{44}$ and $R_{55}$ read respectively:
\ba{a10} \fl &{}& R_{44}\nn\\
\fl &=&\frac{A_{\xi}}{4A}\left(-\frac{2A_{\xi\xi}}{A_{\xi}}+\frac{E_{\xi}}{E}+\frac{A_{\xi}}{A}\right)+\frac{E_t}{4c^2A}
\left(-\frac{2E_{tt}}{E_t}+\frac{A_t}{A}-\frac{B_t}{B}-\frac{C_t}{C}-\frac{D_t}{D}-\frac{F_t}{F}+\frac{E_t}{E}\right)\nn
\\
\fl &+&\frac{B_{\xi}}{4B}\left(-\frac{2B_{\xi\xi}}{B_{\xi}}+\frac{E_{\xi}}{E}+\frac{B_{\xi}}{B}\right)+\frac{E_x}{4B}
\left(-\frac{2E_{xx}}{E_x}+\frac{B_x}{B}-\frac{A_x}{A}-\frac{C_x}{C}-\frac{D_x}{D}-\frac{F_x}{F}+\frac{E_x}{E}\right)\nn
\\
\fl &+&\frac{C_{\xi}}{4C}\left(-\frac{2C_{\xi\xi}}{C_{\xi}}+\frac{E_{\xi}}{E}+\frac{C_{\xi}}{C}\right)+\frac{E_y}{4C}
\left(-\frac{2E_{yy}}{E_y}+\frac{C_y}{C}-\frac{A_y}{A}-\frac{B_y}{B}-\frac{D_y}{D}-\frac{F_y}{F}+\frac{E_y}{E}\right)\nn
\\
\fl &+&\frac{D_{\xi}}{4D}\left(-\frac{2D_{\xi\xi}}{D_{\xi}}+\frac{E_{\xi}}{E}+\frac{D_{\xi}}{D}\right)+\frac{E_z}{4D}
\left(-\frac{2E_{zz}}{E_z}+\frac{D_z}{D}-\frac{A_z}{A}-\frac{B_z}{B}-\frac{C_z}{C}-\frac{F_z}{F}+\frac{E_z}{E}\right)\nn
\\
\fl &+&\frac{F_{\xi}}{4F}\left(-\frac{2F_{\xi\xi}}{F_{\xi}}+\frac{E_{\xi}}{E}+\frac{F_{\xi}}{F}\right)+\frac{E_{\eta}}{4F}
\left(-\frac{2E_{\eta\eta}}{E_{\eta}}+\frac{F_{\eta}}{F}-\frac{A_{\eta}}{A}-\frac{B_{\eta}}{B}-\frac{C_{\eta}}{C}-
\frac{D_{\eta}}{D}+\frac{E_{\eta}}{E}\right)\nn
\\
\fl &\approx&
1 -\frac {\epsilon}{2}
 \left\{(A^{1}-B^{1}-C^{1}-D^{1})_{\xi \xi} -\triangle_3 E^{1}-\frac{E^{1}_{\eta \eta}}{a^2\sin^2\xi} \right.\nn\\
  \fl &-&\left.\frac{1}{a^2\sin^2\xi}\left[ \left( F^1_{\xi \xi}-
  2\frac{\cos2\xi}{\sin2\xi}F^1_\xi\right)
  -\frac{2}{\sin 2\xi}\left( F^1_\xi - 2\frac{\cos\xi}{\sin\xi}F^1 \right)
  -\frac {\sin 2\xi}{2} E^1_{\xi} \right]
   \right\} \, ,
\ea
\ba{a11} \fl &{}& R_{55}\nn\\
\fl &=& \frac{A_{\eta}}{4A}\left(-\frac{2A_{\eta\eta}}{A_{\eta}}+\frac{F_{\eta}}{F}+\frac{A_{\eta}}{A}\right)+\frac{F_t}{4c^2A}
\left(-\frac{2F_{tt}}{F_t}+\frac{A_t}{A}-\frac{B_t}{B}-\frac{C_t}{C}-\frac{D_t}{D}-\frac{E_t}{E}+\frac{F_t}{F}\right)\nn
\\
\fl &+&\frac{B_{\eta}}{4B}\left(-\frac{2B_{\eta\eta}}{B_{\eta}}+\frac{F_{\eta}}{F}+\frac{B_{\eta}}{B}\right)+\frac{F_x}{4B}
\left(-\frac{2F_{xx}}{F_x}+\frac{B_x}{B}-\frac{A_x}{A}-\frac{C_x}{C}-\frac{D_x}{D}-\frac{E_x}{E}+\frac{F_x}{F}\right)\nn
\\
\fl &+&\frac{C_{\eta}}{4C}\left(-\frac{2C_{\eta\eta}}{C_{\eta}}+\frac{F_{\eta}}{F}+\frac{C_{\eta}}{C}\right)+\frac{F_y}{4C}
\left(-\frac{2F_{yy}}{F_y}+\frac{C_y}{C}-\frac{A_y}{A}-\frac{B_y}{B}-\frac{D_y}{D}-\frac{E_y}{E}+\frac{F_y}{F}\right)\nn
\\
\fl &+&\frac{D_{\eta}}{4D}\left(-\frac{2D_{\eta\eta}}{D_{\eta}}+\frac{F_{\eta}}{F}+\frac{D_{\eta}}{D}\right)+\frac{F_z}{4D}
\left(-\frac{2F_{zz}}{F_z}+\frac{D_z}{D}-\frac{A_z}{A}-\frac{B_z}{B}-\frac{C_z}{C}-\frac{E_z}{E}+\frac{F_z}{F}\right)\nn
\\
\fl &+&\frac{E_{\eta}}{4E}\left(-\frac{2E_{\eta\eta}}{E_{\eta}}+\frac{F_{\eta}}{F}+\frac{E_{\eta}}{E}\right)+\frac{F_{\xi}}{4E}
\left(-\frac{2F_{\xi\xi}}{F_{\xi}}+\frac{E_{\xi}}{E}-\frac{A_{\xi}}{A}-\frac{B_{\xi}}{B}-\frac{C_{\xi}}{C}-\frac{D_{\xi}}{D}+\frac{F_{\xi}}{F}\right)\nn
\\
\fl &\approx&
\sin^2\xi-\frac {\epsilon}{2} \left\{(A^{1}-B^{1}-C^{1}-D^{1})_{\eta\eta}-\triangle_3 F^{1}-\frac{E^{1}_{\eta \eta}}{a^2}
-\frac{1}{a^2}\left(  F^1_{\xi\xi}-2\frac{\cos2\xi}{\sin2\xi} F^1_{\xi} \right)\right.\nn
\\
\fl &+&\left. \frac{\sin2\xi}{2} \left( \frac{E^1_\xi}{a^2}+(A^1-B^1-C^1-D^1)_\xi\right)+ \frac{\cos\xi}{a^2\sin\xi}\left( F^1_\xi -
2\frac{\cos\xi}{\sin\xi}F^1\right)\right.\nn
\\
\fl &+&\left. \frac{1}{a^2}\frac{\sin\xi}{\cos\xi}
\left(  F^1_\xi - \sin{2\xi} E^1 \right)
\right\}\, .
\ea

\subsection{Off-diagonal components}

\begin{eqnarray*}
\fl R_{01}&=&-\frac{1}{2C}C_{tx}-\frac{1}{2D}D_{tx}-\frac{1}{2E}E_{tx}-\frac{1}{2F}F_{tx}\\
\fl &+&\frac{A_xC_t}{4AC}+\frac{A_xD_t}{4AD}+\frac{A_xE_t}{4AE}+\frac{A_xF_t}{4AF}
+\frac{B_tC_x}{4BC}+\frac{B_tD_x}{4BD}\\
\fl &+& \frac{B_tE_x}{4BE}+\frac{B_tF_x}{4BF}+\frac{C_tC_x}{4C^2}+\frac{D_tD_x}{4D^2} +\frac{E_tE_x}{4E^2}+\frac{F_tF_x}{4F^2}\, ,
\end{eqnarray*}
\begin{eqnarray*}
\fl R_{02}&=& -\frac{1}{2B}B_{ty}-\frac{1}{2D}D_{ty}-\frac{1}{2E}E_{ty}-\frac{1}{2F}F_{ty}\\
\fl &+& \frac{A_yB_t}{4AB}+\frac{A_yD_t}{4AD}+\frac{A_yE_t}{4AE}+\frac{A_yF_t}{4AF}+
\frac{C_tB_y}{4BC}+\frac{C_tD_y}{4CD}\\
\fl &+&\frac{C_tE_y}{4CE}+\frac{C_tF_y}{4CF}+\frac{B_tB_y}{4B^2}+\frac{D_tD_y}{4D^2}+\frac{E_tE_y}{4E^2}+\frac{F_tF_y}{4F^2}\, ,
\end{eqnarray*}
\begin{eqnarray*}
\fl R_{03}&=&-\frac{1}{2B}B_{tz}-\frac{1}{2C}C_{tz}-\frac{1}{2E}E_{tz}-\frac{1}{2F}F_{tz}\\
\fl &+&\frac{A_zB_t}{4AB}+\frac{A_zC_t}{4AC}+\frac{A_zE_t}{4AE}+\frac{A_zF_t}{4AF}+
\frac{D_tB_z}{4BD}+\frac{D_tC_z}{4CD}\\
\fl &+&\frac{D_tE_z}{4DE}+\frac{D_tF_z}{4DF}+\frac{B_tB_z}{4B^2}+\frac{C_tC_z}{4C^2}+\frac{E_tE_z}{4E^2}+\frac{F_tF_z}{4F^2}\, ,
\end{eqnarray*}
\begin{eqnarray*}
\fl R_{04}&=&-\frac{1}{2B}B_{t\xi}-\frac{1}{2C}C_{t\xi}-\frac{1}{2D}D_{t\xi}-\frac{1}{2F}F_{t\xi}\\
\fl &+&\frac{A_{\xi}B_t}{4AB}+\frac{A_{\xi}C_t}{4AC}+\frac{A_{\xi}D_t}{4AD}+\frac{A_{\xi}F_t}{4AF}+
\frac{E_tB_{\xi}}{4BE}+\frac{E_tC_{\xi}}{4EC}\\
\fl &+& \frac{E_tD_{\xi}}{4ED}+\frac{E_tF_{\xi}}{4EF}+
\frac{B_tB_{\xi}}{4B^2}+\frac{C_tC_{\xi}}{4C^2}+\frac{D_tD_{\xi}}{4D^2}+\frac{F_tF_{\xi}}{4F^2}\, ,
\end{eqnarray*}
\begin{eqnarray*}
\fl R_{05}&=&-\frac{1}{2B}B_{t\eta}-\frac{1}{2C}C_{t\eta}-\frac{1}{2D}D_{t\eta}-\frac{1}{2E}E_{t\eta}\\
\fl &+& \frac{A_{\eta}B_t}{4AB}+\frac{A_{\eta}C_t}{4AC}+\frac{A_{\eta}D_t}{4AD}+\frac{A_{\eta}E_t}{4AE}+
\frac{F_tB_{\eta}}{4BF}+\frac{F_tC_{\eta}}{4FC}\\
\fl &+& \frac{F_tD_{\eta}}{4FD}+\frac{F_tE_{\eta}}{4FE}+
\frac{B_tB_{\eta}}{4B^2}+\frac{C_tC_{\eta}}{4C^2}+\frac{D_tD_{\eta}}{4D^2}+\frac{E_tE_{\eta}}{4E^2}\, .
\end{eqnarray*}
Obviously, for the static metrics these components are identically equal to zero. Let us now calculate the remaining $10$ off-diagonal components:
\ba{a12}
\fl R_{12}&=&-\frac{1}{2A}A_{xy}-\frac{1}{2D}D_{xy}-\frac{1}{2E}E_{xy}-\frac{1}{2F}F_{xy}\nn\\
\fl &+&\frac{B_yA_x}{4AB}+\frac{B_yD_x}{4BD}+\frac{B_yE_x}{4BE}+\frac{B_yF_x}{4BF}+\frac{C_xA_y}{4AC}+\frac{C_xD_y}{4CD}\nn\\
\fl &+& \frac{C_xE_y}{4CE}+\frac{C_xF_y}{4CF}+
\frac{A_xA_y}{4A^2}+\frac{D_xD_y}{4D^2}+\frac{E_xE_y}{4E^2}+\frac{F_xF_y}{4F^2}\nn\\
\fl &\approx&
\epsilon\left(-\frac{1}{2}A^1+\frac{1}{2}D^1+\frac{1}{2a^2}E^1+\frac{1}{2a^2\sin^2\xi}F^1\right)_{xy}\, ,
\ea
\ba{a13}
\fl R_{13}&=&-\frac{1}{2A}A_{xz}-\frac{1}{2C}C_{xz}-\frac{1}{2E}E_{xz}-\frac{1}{2F}F_{xz}\nn \\
\fl &+& \frac{B_zA_x}{4AB}+\frac{B_zC_x}{4BC}+\frac{B_zE_x}{4BE}+\frac{B_zF_x}{4BF}+\frac{D_xA_z}{4AD}+\frac{D_xC_z}{4CD}\nn \\
\fl &+& \frac{D_xE_z}{4DE}+\frac{D_xF_z}{4DF}+
\frac{A_xA_z}{4A^2}+\frac{C_xC_z}{4C^2}+\frac{E_xE_z}{4E^2}+\frac{F_xF_z}{4F^2}\nn \\
\fl &\approx&
\epsilon\left(-\frac{1}{2}A^1+\frac{1}{2}C^1+\frac{1}{2a^2}E^1+\frac{1}{2a^2\sin^2\xi}F^1\right)_{xz}\, ,
\ea
\ba{a14}
\fl R_{23}&=&-\frac{1}{2A}A_{yz}-\frac{1}{2B}B_{yz}-\frac{1}{2E}E_{yz}-\frac{1}{2F}F_{yz}\nn \\
\fl &+& \frac{C_zA_y}{4AC}+\frac{C_zB_y}{4BC}+\frac{C_zE_y}{4CE}+\frac{C_zF_y}{4CF}+\frac{D_yA_z}{4AD}+\frac{D_yB_z}{4BD}\nn \\
\fl &+& \frac{D_yE_z}{4DE}+\frac{D_yF_z}{4DF}+
\frac{A_yA_z}{4A^2}+\frac{B_yB_z}{4B^2}+\frac{E_yE_z}{4E^2}+\frac{F_yF_z}{4F^2}\nn \\
\fl &\approx&
\epsilon\left(-\frac{1}{2}A^1+\frac{1}{2}B^1+\frac{1}{2a^2}E^1+\frac{1}{2a^2\sin^2\xi}F^1\right)_{yz}\, ,
\ea
\ba{a15}
\fl R_{15}&=&-\frac{1}{2A}A_{x\eta}-\frac{1}{2C}C_{x\eta}-\frac{1}{2D}D_{x\eta}-\frac{1}{2E}E_{x\eta}\nn \\
\fl &+& \frac{B_{\eta}A_x}{4AB}+\frac{B_{\eta}C_x}{4BC}+\frac{B_{\eta}D_x}{4BD}+\frac{B_{\eta}E_x}{4BE}+
\frac{F_xA_{\eta}}{4AF}+\frac{F_xC_{\eta}}{4FC}\nn \\
\fl &+& \frac{F_xD_{\eta}}{4FD}+\frac{F_xE_{\eta}}{4FE}+
\frac{A_xA_{\eta}}{4A^2}+\frac{C_xC_{\eta}}{4C^2}+\frac{D_xD_{\eta}}{4D^2}+\frac{E_xE_{\eta}}{4E^2}\nn \\
\fl &\approx&
\epsilon\left(-\frac{1}{2}A^1+\frac{1}{2}C^1+\frac{1}{2}D^1+\frac{1}{2a^2}E^1\right)_{x\eta}\, ,
\ea
\ba{a16}
\fl R_{25}&=&-\frac{1}{2A}A_{y\eta}-\frac{1}{2B}B_{y\eta}-\frac{1}{2D}D_{y\eta}-\frac{1}{2E}E_{y\eta}\nn \\
\fl &+&\frac{C_{\eta}A_y}{4AC}+\frac{C_{\eta}B_y}{4BC}+\frac{C_{\eta}D_y}{4CD}+\frac{C_{\eta}E_y}{4CE}+
\frac{F_yA_{\eta}}{4AF}+\frac{F_yB_{\eta}}{4BF}\nn \\
\fl &+& \frac{F_yD_{\eta}}{4FD}+\frac{F_yE_{\eta}}{4FE}
+\frac{A_yA_{\eta}}{4A^2}+\frac{B_yB_{\eta}}{4B^2}+\frac{D_yD_{\eta}}{4D^2}+\frac{E_yE_{\eta}}{4E^2}\nn \\
\fl &\approx&
\epsilon\left(-\frac{1}{2}A^1+\frac{1}{2}B^1+\frac{1}{2}D^1+\frac{1}{2a^2}E^1\right)_{y\eta}\, ,
\ea
\ba{a17}
\fl R_{35}&=&-\frac{1}{2A}A_{z\eta}-\frac{1}{2B}B_{z\eta}-\frac{1}{2C}C_{z\eta}-\frac{1}{2E}E_{z\eta}\nn \\
\fl &+& \frac{D_{\eta}A_z}{4AD}+\frac{D_{\eta}B_z}{4BD}+\frac{D_{\eta}C_z}{4CD}+\frac{D_{\eta}E_z}{4ED}+
\frac{F_zA_{\eta}}{4AF}+\frac{F_zB_{\eta}}{4BF}\nn \\
\fl &+& \frac{F_zC_{\eta}}{4CF}+\frac{F_zE_{\eta}}{4FE}+
\frac{A_zA_{\eta}}{4A^2}+\frac{B_zB_{\eta}}{4B^2}+\frac{C_zC_{\eta}}{4C^2}+\frac{E_zE_{\eta}}{4E^2}\nn \\
\fl &\approx&
\epsilon\left(-\frac{1}{2}A^1+\frac{1}{2}B^1+\frac{1}{2}C^1+\frac{1}{2a^2}E^1\right)_{z\eta}\, ,
\ea
\ba{a18}
\fl R_{14}&=& -\frac{1}{2A}A_{x\xi}-\frac{1}{2C}C_{x\xi}-\frac{1}{2D}D_{x\xi}-\frac{1}{2F}F_{x\xi}\nn \\
\fl &+& \frac{B_{\xi}A_x}{4AB}+\frac{B_{\xi}C_x}{4BC}+\frac{B_{\xi}D_x}{4BD}+\frac{B_{\xi}F_x}{4BF}+
\frac{E_xA_{\xi}}{4AE}+\frac{E_xC_{\xi}}{4EC}\nn \\
\fl &+& \frac{E_xD_{\xi}}{4ED}+\frac{E_xF_{\xi}}{4EF}+
\frac{A_xA_{\xi}}{4A^2}+\frac{C_xC_{\xi}}{4C^2}+\frac{D_xD_{\xi}}{4D^2}+\frac{F_xF_{\xi}}{4F^2}\nn \\
\fl &\approx& \epsilon\left(\frac{1}{2}( -A^1+C^1+D^1)_{\xi}+
\frac{1}{2a^2\sin^2\xi}F_{\xi}^1-\frac{\cos\xi}{2a^2\sin\xi}E^1-\frac{\cos\xi}{2a^2\sin^3\xi}F^1\right)_x\, ,
\ea
\ba{a19}
\fl R_{24}&=&-\frac{1}{2A}A_{y\xi}-\frac{1}{2B}B_{y\xi}-\frac{1}{2D}D_{y\xi}-\frac{1}{2F}F_{y\xi}\nn \\
\fl &+& \frac{C_{\xi}A_y}{4AC}+\frac{C_{\xi}B_y}{4BC}+\frac{C_{\xi}D_y}{4CD}+\frac{C_{\xi}F_y}{4CF}+
\frac{E_yA_{\xi}}{4AE}+\frac{E_yB_{\xi}}{4BE}\nn \\
\fl &+& \frac{E_yD_{\xi}}{4ED}+\frac{E_yF_{\xi}}{4EF}+
\frac{A_yA_{\xi}}{4A^2}+\frac{B_yB_{\xi}}{4B^2}+\frac{D_yD_{\xi}}{4D^2}+\frac{F_yF_{\xi}}{4F^2}\nn \\
\fl &\approx& \epsilon\left(\frac{1}{2}(-A^1+B^1+D^1)_{\xi}+\frac{1}{2a^2\sin^2\xi}F_{\xi}^1-
\frac{\cos\xi}{2a^2\sin\xi}E^1-\frac{\cos\xi}{2a^2\sin^3\xi}F^1\right)_y\, ,
\ea
\ba{a20}
\fl R_{34}&=&-\frac{1}{2A}A_{z\xi}-\frac{1}{2B}B_{z\xi}-\frac{1}{2C}C_{z\xi}-\frac{1}{2F}F_{z\xi}\nn \\
\fl &+& \frac{D_{\xi}A_z}{4AD}+\frac{D_{\xi}B_z}{4BD}+\frac{D_{\xi}C_z}{4CD}+\frac{D_{\xi}F_z}{4FD}
+\frac{E_zA_{\xi}}{4AE}+\frac{E_zB_{\xi}}{4BE}\nn \\
\fl &+& \frac{E_zC_{\xi}}{4CE}+\frac{E_zF_{\xi}}{4FE}
+\frac{A_zA_{\xi}}{4A^2}+\frac{B_zB_{\xi}}{4B^2}+\frac{C_zC_{\xi}}{4C^2}+\frac{F_zF_{\xi}}{4F^2}\nn \\
\fl &\approx& \epsilon\left(\frac{1}{2}(-A^1+B^1+C^1)_{\xi}+\frac{1}{2a^2\sin^2\xi}F_{\xi}^1-
\frac{\cos\xi}{2a^2\sin\xi}E^1-\frac{\cos\xi}{2a^2\sin^3\xi}F^1\right)_z\, ,
\ea
\ba{a21}
\fl R_{45}&=& -\frac{1}{2A}A_{\xi\eta}-\frac{1}{2B}B_{\xi\eta}-\frac{1}{2C}C_{\xi\eta}-\frac{1}{2D}D_{\xi\eta}\nn \\
\fl &+& \frac{E_{\eta}A_\xi}{4AE}+\frac{E_{\eta}B_\xi}{4BE}+\frac{E_{\eta}C_\xi}{4CE}+\frac{E_{\eta}D_\xi}{4DE}
+\frac{F_\xi A_{\eta}}{4AF}+\frac{F_\xi B_{\eta}}{4BF}\nn \\
\fl &+& \frac{F_\xi C_{\eta}}{4CF}+\frac{F_\xi D_{\eta}}{4FD}+\frac{A_\xi A_{\eta}}{4A^2}+\frac{B_\xi B_{\eta}}{4B^2}+\frac{C_\xi C_{\eta}}{4C^2}+\frac{D_\xi
D_{\eta}}{4D^2}\nn \\
\fl &\approx&\epsilon\left(\frac{1}{2}(-A^1+B^1+C^1+D^1)_{\xi}+
\frac{\cos\xi}{2\sin\xi}(A^1-B^1-C^1-D^1)\right)_{\eta}\, .
\ea

\section{Appendix B: Relations between metric coefficients}
\renewcommand{\theequation}{B\arabic{equation}}
\setcounter{equation}{0}

First, we investigate expressions \rf{a12}-\rf{a14} in the case $R_{12}=R_{13}=R_{23}=0$. It can be easily seen that the equation $R_{12}=0$ has a solution
$$
-\frac{1}{2}A^1+\frac{1}{2}D^1+\frac{1}{2a^2}E^1+\frac{1}{2a^2\sin^2\xi}F^1=C_1(z,\xi,\eta)f_1(x)+C_2(z,\xi,\eta)f_2(y)\, ,
$$
where $C_1(z,\xi,\eta), C_2(z,\xi,\eta), f_1(x)$ and $f_2(y)$ are arbitrary functions. We also assume that in the limit $|x|,|y|,|z|\to +\infty$ the perturbed metrics
reduces to the background one. Thus, all perturbations $A^1,B^1,C^1,D^1,E^1$ and $F^1$ as well as their partial derivatives vanish in this limit. Therefore, the right
hand side of the above equation is equal to zero. Similar reasoning can be applied to equations $R_{13}=0$ and $R_{23}=0$. Then, we arrive at the following relations:
\be{b1} B^1=C^1=D^1=A^1-\frac{1}{a^2}E^1-\frac{1}{a^2\sin^2\xi}F^1\, . \ee

We consider models where the Einstein equation for all off-diagonal components is reduced to
\be{b2}
R_{ik}=0 \quad \mbox{for}\quad i\neq k\, .
\ee
We want to analyze these equations for components \rf{a15}-\rf{a21} with regard to the relations \rf{b1}.

First, it can be easily seen that Einstein equations \rf{b2} for components \rf{a18}-\rf{a20} give
\be{b3}
\fl -A_{\xi}^1+B_{\xi}^1+C_{\xi}^1+\frac{1}{a^2\sin^2\xi}F_{\xi}^1-\frac{\cos\xi}{a^2\sin\xi}E^1-\frac{\cos\xi}{a^2\sin^3\xi}F^1=C_3(\xi,\eta)\, ,
\ee
where $C_3(\xi,\eta)$ is an arbitrary function. From the boundary conditions at $|x|,|y|,|z|\to +\infty$ we find that $C_3(\xi,\eta)=0$. Taking it into account, we get
from \rf{b1} and \rf{b3} respectively
\be{b4}
-A^1+B^1+\frac{1}{a^2\sin^2\xi}F^1=-\frac{1}{a^2}E^1
\ee
and
\be{b5}
-A_{\xi}^1+B_{\xi}^1+\frac{1}{a^2\sin^2\xi}F_{\xi}^1-\frac{\cos\xi}{a^2\sin\xi}E^1-\frac{\cos\xi}{a^2\sin^3\xi}F^1=-B_{\xi}^1\, .
\ee
Differentiating \rf{b4} with respect to $\xi$, we obtain
\be{b6}
-A^1_{\xi}+B^1_{\xi}+\frac{1}{a^2\sin^2\xi}F^1_{\xi}-\frac{2\cos\xi}{a^2\sin^3\xi}F^1=-\frac{1}{a^2}E^1_{\xi}\, .
\ee
Subtraction \rf{b6} from \rf{b5} yields
\be{b7}
\frac{1}{a^2\sin^2\xi}F^1 =-B_{\xi}^1\tan\xi+\frac{1}{a^2}E^1_{\xi}\tan\xi+\frac{1}{a^2}E^1\, .
\ee

Let us investigate two separate cases.

\vspace{0.3cm}

{\it{1. Smeared extra dimensions}}

\vspace{0.3cm}

First, we consider the matter source which is uniformly smeared over the internal space. It results in the metric coefficients $A^1,B^1,C^1,D^1$ and $E^1$  depending
only on the external coordinates $x,y$ and $z$ \cite{EZ2}. We do not require that the diagonal Einstein equations have the form \rf{2.13}-\rf{2.16}, but the off-diagonal
components must be like \rf{b2}. Then, equations $R_{15}=R_{25}=R_{35}=R_{45}=0$ (where these off-diagonal components are defined by \rf{a15}-\rf{a17},\rf{a21}) are
automatically satisfied. It can be easily seen from \rf{b1} that the coefficient $F^1 \sim \sin^2\xi$. Moreover, to satisfy the equation \rf{b7}, it should have the form
\be{b8}
F^1=E^1\sin^2\xi\, .
\ee
Therefore, \rf{b1} can be rewritten in the form
\be{b9}
-A^1 + B^1 +\frac{2}{a^2}E^1=0\, .
\ee

\vspace{0.3cm}

{\it{2. Arbitrary rest mass density $\rho(\bf{r_5})$}}

\vspace{0.3cm}

In this case, the rest mass density $\rho(\bf{r_5})$ of a compact astrophysical object is an arbitrary function of all five spatial coordinates. Here, as a particular
example, the matter source can also be smeared over the internal space. However, the main difference from the previous case is that the diagonal Einstein equations
should have the form \rf{2.13}-\rf{2.16}. This leads to additional conditions
\be{b10}
B^1 =C^1=D^1=\frac{1}{3}A^1\, ,
\ee
which follows from equations \rf{2.17}, \rf{2.18} and \rf{b1}. The case 1 is related to the models where \rf{b1} is still valid but the relation $B^1=A^1/3$ may be
violated. Therefore, taking into account relations \rf{b10}, the Einstein equation $R_{45}=0$ for \rf{a21} is automatically satisfied. Let us consider equations \rf{b2}
for components \rf{a18}-\rf{a20}. Substitution \rf{b7} back into \rf{b4} gives
\be{b11}
-2B^1 + \frac{2}{a^2}E^1=B_{\xi}^1\tan\xi-\frac{1}{a^2}E^1_{\xi}\tan\xi\, ,
\ee
where we take into account the relation \rf{b10}. We seek the solution of this equation in the form
\be{b12}
E^1 = a^2B^1+\tilde{E}^1,\quad 2\tilde{E}^1=-\tilde{E}^1_{\xi}\tan\xi\, .
\ee
The solution of the latter equation is $\tilde{E}^1=C_4({\bf r}_3,\eta)/\sin^2\xi$, where $C_4({\bf r}_3,\eta)$ is an arbitrary function. The function $\tilde{E}^1$
diverges when $\xi\rightarrow 0,\pi$. To avoid this problem, we require that $C_4({\bf r}_3,\eta)=0$. Thus, $\tilde{E}^1=0$ and $E^1=a^2B^1$. In turn, from the equation
\rf{b7}, we obtain: $F^1=E^1\sin^2\xi=a^2B^1\sin^2\xi$. Thus, perturbations of the metric coefficients are related as follows:
\be{b13}
B^1=C^1=D^1=A^1/3,\quad E^1=a^2B^1,\quad F^1=E^1\sin^2\xi\, .
\ee
Therefore, the equality \rf{b9} holds also in this case. Taking into account these relations, we can easily verify that Einstein equations \rf{b2} for components
\rf{a15}-\rf{a17} are automatically satisfied. Denoting $A^1\equiv 2\varphi/c^2$, we can conclude that the perturbed metrics should have the following structure:
\ba{b14}
ds^2&\approx& \left(1+\frac{2\varphi}{c^2}\right)c^2dt^2+\left(-1+\frac{2\varphi}{3c^2}\right)
\left(dx^2+dy^2+dz^2\right)\nn \\
&+&a^2\left(-1+\frac{2\varphi}{3c^2}\right)\left(d\xi^2+\sin^2\xi d\eta^2\right)\, .
\ea

To complete this appendix (the case 2), we consider now $44$ and $55$ Ricci tensor components \rf{a10} and \rf{a11}. With the help of relations \rf{b13}, it is not
difficult to verify that
\be{b15}
\sin^2\xi \times R_{44} = R_{55}\, .
\ee
Moreover, the expression \rf{a10} can be rewritten in the following form:
\be{b16}
R_{44} \approx 1 +\frac{\epsilon}{2}\left[\triangle_3 E^1 +\frac{1}{a^2}\triangle_{\xi\eta}E^1\right]=1 + \frac{a^2}{3}R_{00}\, .
\ee
Similarly, the components $11$ and $22$ are connected with the component $00$ as follows:
\be{b17}
R_{11}=R_{22}=R_{33}=\frac{1}{3}R_{00}\, .
\ee

\end{document}